# Solar Wind Reflected Ion Properties at Earth's Bow Shock: Dependence on Upstream Conditions and Shock Geometry

**Runyi Liu[1], Terry Liu[2], Kun Zhang[3], Vassilis Angelopoulos[1], Siqi Zhao[4]**

[1]Department of Earth, Planetary, and Space Sciences, University of California, Los Angeles, Los Angeles, CA, USA
[2]Shandong Key Laboratory of Space Environment and Exploration Technology, Institute of Space Sciences, School of Space Science and Technology, Shandong University, Shandong, China.
[3]Department of Physical Sciences, Embry-Riddle Aeronautical University, Daytona Beach, FL, 32114
[4]Institute of Physics and Astronomy, University of Potsdam, D-14476 Potsdam, Germany

**Key Points:**

- Ion reflection ratio at Earth's bow shock decreases with $\theta_{Bn}$ and increases with magnetic compression, controlled by shock geometry.
- Reflected ion energies deviate from idealized models; a combined adiabatic–specular representation improves agreement with observations.
- Reflected ion temperature correlates with upstream magnetic field and solar wind temperature, dominated by magnetic fluctuations.

Corresponding author: Runyi Liu and Terry Liu, `runyiliu11@ucla.edu`; `terryliuzixu@ucla.edu`

**Abstract**

Solar wind ion reflection at collisionless shocks regulates foreshock plasma dynamics, yet the quantitative dependence of reflected ion properties on upstream and shock-related parameters remains unclear, causing difficulties in predicting foreshock disturbances. We present a statistical study of solar wind reflected ions near the Earth's bow shock using THEMIS observations from 59 well-defined shock crossings between 2016 and 2019. Reflected ion moments are derived after removal of the solar wind core and compared with upstream and shock parameters. The reflection ratio decreases with increasing angle between interplanetary magnetic field and shock normal, $\theta_{Bn}$, and increases with magnetic compression ratio, indicating that shock geometry and magnetic compression primarily regulate ion reflection. Reflected ion energies deviate from individual idealized reflection models: the adiabatic model overestimates total ion energy, whereas the specular model captures the perpendicular component. A combined adiabatic–specular representation improves the linear energy correspondence, and model–observation agreement increases under quasi-perpendicular shock conditions for all comparisons. Reflected ion temperature correlates with upstream magnetic field strength and solar wind temperature, and shows a substantially stronger dependence on magnetic field fluctuation energy. Overall, reflected ion properties are primarily governed by upstream conditions and shock structure, with magnetic field fluctuations contributing to ion thermalization, providing observational constraints on ion reflection and heating at Earth's bow shock.

# 1 Introduction

The terrestrial ion foreshock is the region upstream of the Earth's bow shock populated by backstreaming ions from the bow shock (Eastwood et al., 2005). These foreshock ions excite ultra–low-frequency (ULF) waves (Greenstadt et al., 1995) and directly control the formation of foreshock transients (Liu et al., 2022), thereby strongly influencing upstream plasma dynamics and solar wind-magnetosphere coupling.

Foreshock ions originate primarily from two sources. One source is the reflection of incident solar wind ions at the bow shock (e.g., Gosling et al., 1978), which produces backstreaming ion populations propagating upstream primarily along the interplanetary magnetic field (IMF) (e.g., Bonifazi & Moreno, 1981). The other source is magnetosheath ion leakage, where downstream ions escape across the bow shock and access the upstream region (Edmiston et al., 1982). While both mechanisms contribute to the foreshock ion population, magnetosheath leakage is influenced by both upstream parameters and downstream ion distributions (Liu, Angelopoulos, Zhang, et al., 2024). In contrast, reflected solar wind ions are directly controlled by upstream parameters and the shock structure itself (Eastwood et al., 2005).

Observations show that the reflection ratio of solar wind ions at the Earth's bow shock tends to increase with the solar wind Mach number and can reach values of about 15–25% under quasi-perpendicular conditions (Sckopke et al., 1983). To explain how ions are reflected at the shock, two primary reflection models have been proposed. In the specular reflection model, ions are reflected by the cross-shock electrostatic potential, resulting in a reversal of the ion velocity component normal to the shock surface while preserving the tangential components (Kennel et al., 2013). Alternatively, in the adiabatic reflection model, ion reflection arises from conservation of the first adiabatic invariant as ions encounter the enhanced magnetic field strength within the shock ramp, leading to reflection due to magnetic mirroring (Schwartz et al., 1983).

However, previous observational studies have shown that foreshock ion velocity distributions often exhibit features that cannot be fully explained by either specular or adiabatic reflection alone (e.g., Liu, Angelopoulos, & Hietala, 2017; Meziane et al., 2007). Rather than being mutually exclusive, these two mechanisms may operate concurrently. Motivated by this perspective, hybrid reflection scenarios have been proposed in which both specular and adiabatic processes contribute to solar wind ion reflection (Sharma & Gedalin, 2023), although direct observational validation remains limited. Moreover, existing models primarily focus on reflected ion bulk velocities and do not provide quantitative predictions for other moments, such as the density (or the ratio to the solar wind ions) or the temperature of the reflected ion population.

In addition, an observational study has attempted to examine statistical relationships between foreshock ion properties and upstream solar wind conditions (Liu et al., 2022). However, their analyses generally considered mixed foreshock ion populations and relied on bow shock models to estimate the spacecraft distance from the shock with large uncertainties, so that contributions from magnetosheath leakage ions and spatial evolution effects may be included in the observed ion properties.

In this study, we statistically quantify the dependence of reflected ion properties on upstream and shock-related parameters near the Earth's bow shock in order to minimize the role of distance to the bow shock. Specifically, we investigate the reflection ratio and temperature of reflected ions and evaluate their relationships with upstream solar wind conditions and shock geometry. In addition, we assess the consistency between observed reflected ion velocities and predictions from specular and adiabatic reflection models. This work provides observational constraints on the multi-parameter control of reflected ion properties at Earth's bow shock. Additionally, several studies (Liu, Vu, Zhang, et al., 2023; Liu, Angelopoulos, et al., 2023; Liu, Vu, Angelopoulos, & Zhang, 2023) have developed a predictive model of foreshock transient disturbances, but foreshock ion parameters remain required as inputs. Establishing solar wind-foreshock ion relationships improves the predictability of foreshock ion characteristics and contributes to a better understanding of foreshock-driven space weather phenomena in the near-Earth environment.

## 2 Data & Methodology

This study uses in situ observations from the Time History of Events and Macroscale Interactions during Substorms (THEMIS) mission (Angelopoulos, 2008) to investigate foreshock ion properties upstream of Earth's bow shock. THEMIS provides continuous plasma and magnetic field measurements in the near-Earth solar wind over multiple years, enabling a statistical analysis of bow shock crossing events under a wide range of upstream conditions. Ion measurements are obtained from the Electrostatic Analyzer (ESA) (McFadden et al., 2008), which resolves three-dimensional ion velocity distribution functions (VDFs). From the ion VDFs, the number density, bulk velocity, and temperature of ions are calculated and used as the primary moments in this study. Magnetic field measurements are obtained from the Fluxgate Magnetometer (FGM) (Auster et al., 2008).

We examine bow shock crossings observed between 2016 and 2019 using event list from Liu, Angelopoulos, Zhang, et al. (2024). Bow shock crossing events are selected according to the following criteria: (1) each event contains a single, well-defined bow shock crossing; (2) upstream and downstream intervals exhibit no large-scale variations in plasma and magnetic field parameters that would prevent the determination of representative upstream and downstream conditions. These two criteria are imposed to ensure that the assumptions required for the determination of bow shock parameters are satisfied. Applying these criteria, a total of 59 bow shock crossing events are selected for statistical analysis; a complete list of event times and derived shock parameters is provided in Table S1 of the Supporting Information.

For each selected event, the bow shock is treated as a planar structure over the analysis interval. Under this assumption, the shock normal direction is determined using a mixed-mode method that combines changes in plasma flow velocity, $\Delta V$, and magnetic field, $\Delta B$, across the shock (Paschmann & Daly, 1998). Upstream plasma flow velocity used in the mixed-mode analysis is obtained from the OMNI solar wind data, while upstream magnetic field measurements are taken from THEMIS FGM. Downstream plasma velocity and magnetic field measurements are derived from THEMIS ESA and FGM, respectively. The bow shock is further assumed to propagate with a constant velocity during each event. Under this assumption, the shock velocity is determined using mass conservation across the shock. Plasma density and bulk velocity on both sides of the shock are obtained from THEMIS ESA measurements and are averaged over the selected upstream and downstream intervals. For each event, the shock normal direction and shock velocity are determined using plasma and magnetic field quantities averaged over the same upstream and downstream time intervals. Based on the derived shock normal direction and shock velocity, plasma velocities are transformed from the spacecraft frame into the normal incidence frame (NIF). All velocity-related quantities discussed in this study are evaluated in the NIF, which provides a consistent reference frame tied to the local shock geometry. In addition to the shock normal direction and shock velocity, the magnetic compression ratio across the bow shock is also evaluated for each event. Two definitions of the magnetic compression ratio are considered. The first is defined as the ratio of the downstream average magnetic field magnitude to the upstream average magnetic field magnitude. The second is defined using the maximum downstream magnetic field magnitude associated with the overshoot, normalized by the upstream average magnetic field magnitude. To quantify magnetic field fluctuations, we compute $\mathrm{var}(\mathbf{B})$ for each event as the variance of the magnetic field vector relative to its event-averaged mean. Only events containing at least ten spin-averaged magnetic field measurements are included to ensure statistical robustness.

Figure 1 shows a representative bow shock crossing event observed by THEMIS-D (THD), illustrating upstream ion populations and shock-related parameters prior to detailed data preprocessing. Figures 1a–d present the in situ magnetic field and plasma observations: (a) magnetic field components and magnitude in GSE coordinates, (b) ion number density, (c) ion bulk velocity, and (d) ion energy–flux spectrum. In Figure 1d, the red dashed vertical line marks the bow shock crossing time, identified as the moment when the solar wind ion core disappears in the ion energy–flux spectrum. Immediately after the crossing, the ion energy–flux spectrum exhibits enhanced high-energy fluxes characteristic of magnetosheath ion leakage, which are clearly distinguishable from the more extended upstream foreshock ion population observed farther upstream. Figure 1e displays the shock normal angle $\theta_{Bn}$. For comparisons with the adiabatic reflection model, events with near-perpendicular geometries are excluded because the model prediction becomes unrealistic as $\theta_{Bn}$ approaches 90°.

The following data preprocessing steps are applied sequentially to obtain the foreshock ion measurements used in this study. First, the solar wind ion core is identified and removed from the ion VDFs to isolate the foreshock ion population. The solar wind ion core is located as the peak energy–flux bin in the ion VDF, and all measurements within a velocity-space radius centered on this peak are removed. The radius is initially set to the solar wind bulk speed to separate solar wind and foreshock ion populations (Liu, Angelopoulos, Hietala, & Wilson III, 2017). In cases where the solar wind ions are significantly thermalized and the default radius does not fully remove the solar wind ion, the resulting ion VDFs are visually inspected and the velocity-space radius is adjusted to ensure complete removal of the solar wind ion core (Liu et al., 2022). The density associated with the identified solar wind ion core is retained and used as the solar wind density in subsequent analyses. Ion moments (Figures 1f-h) are then calculated from the remaining VDFs and retained at the spacecraft spin-averaged cadence using ESA reduced-mode measurements. In the following, these measurements are referred to as spin-averaged data to distinguish them from quantities aggregated at the event level. Second, time intervals associated with foreshock transients, as well as intervals corresponding to solar wind discontinuities and the periods following them, are manually excluded. Third, a density ratio threshold is applied by requiring the foreshock density to be less than 0.2 times the solar wind density in order to exclude intervals likely dominated by magnetosheath ion leakage. This value is motivated by the expected upper limit of the solar wind ion reflection ratio at the bow shock, which is typically on the order of ∼20% (e.g., Sckopke et al., 1983). Fourth, for spin-averaged data within 150 seconds upstream of the bow shock crossing, the geometric mean foreshock ion density (i.e., the mean computed in logarithmic space and then converted back to linear space) is used to represent the typical local upstream level for that event. Spin-averaged data whose densities exceed 120 % of this value (in linear space) are removed as outliers in order to further suppress residual magnetosheath ion leakage. Fifth, data points with foreshock ion number densities below 0.05 cm$^{-3}$ are excluded due to ESA noise level (Liu et al., 2022). Finally, only data points within 90 seconds upstream of the bow shock crossing are selected in order to minimize the effects of spatial evolution of foreshock ions with increasing distance from the shock. The analysis interval in Figure 1 is an example satisfying the above criteria.

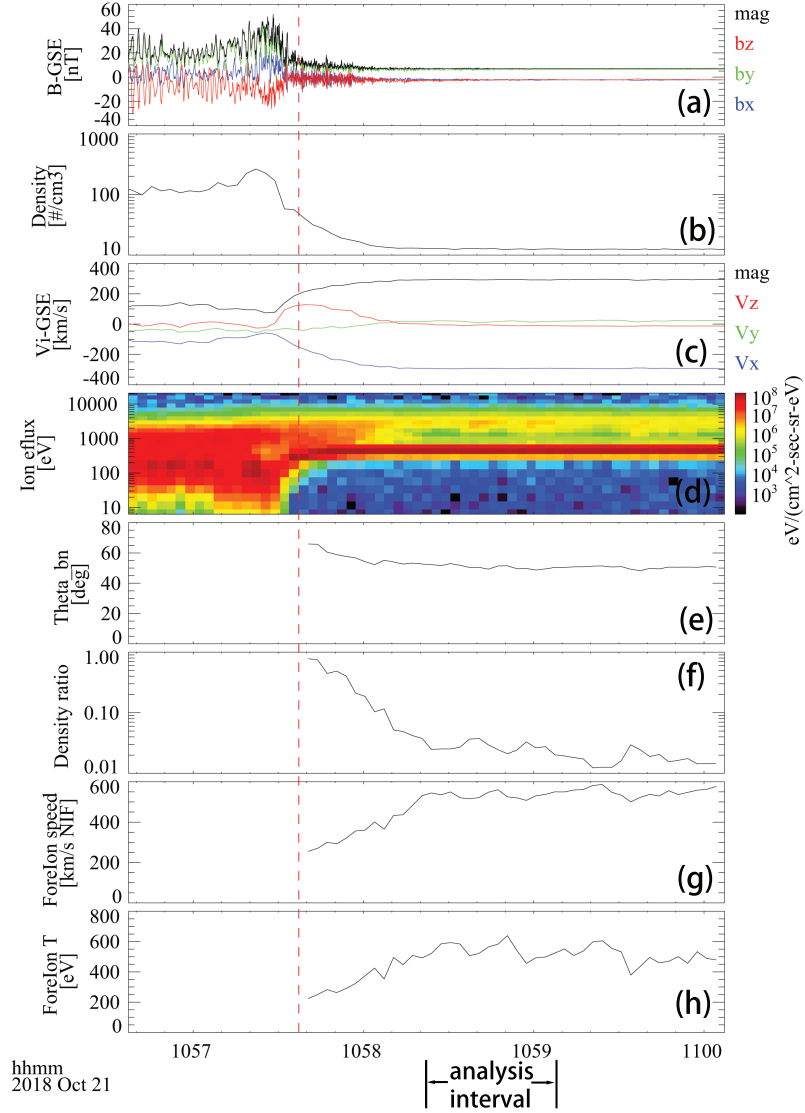

**Figure 1.** Representative bow shock crossing event observed by THEMIS-D (THD) on 21 October 2018. (a) Magnetic field components and magnitude in GSE coordinates measured by FGM. (b) Ion number density derived from ESA measurements. (c) Ion bulk velocity components and magnitude in GSE coordinates. (d) Ion energy–flux spectrum. The red vertical dashed line marks the identified bow shock crossing time. (e) Shock geometry parameter $\theta_{Bn}$, defined as the angle between the upstream magnetic field and the shock normal. (f)–(h) Foreshock ion moments after removal of the solar wind ion core: (f) density ratio (foreshock ion density normalized to solar wind density), (g) foreshock ion bulk speed in the normal incidence frame (NIF), and (h) foreshock ion temperature. The horizontal bar below the time axis marks the upstream interval retained for statistical analysis after the data preprocessing steps described in Section 2.

In addition to the spin-averaged data (Figures 1f-h) described above, event-level quantities are also constructed for statistical analysis. For each bow shock crossing event, upstream parameters, shock-related parameters, and foreshock ion moments are aggregated from the corresponding spin-averaged data points using either the mean or the median, depending on the analysis.

To quantify the relationships between reflected ion moments and upstream or shock-related parameters, we employ Pearson and Spearman correlation coefficients to assess linear and monotonic dependencies, respectively. Because ion moments in the solar wind and foreshock region can be influenced by multiple interrelated parameters, partial correlation analysis is additionally performed to isolate the direct association between a given parameter and the ion moments of interest while controlling for selected variables. In the following figures, Pearson correlation coefficients are denoted by $r$, Spearman rank correlation coefficients by $\rho$, and their corresponding partial correlation coefficients by $pr$ and $p\rho$. Correlation coefficients are considered statistically significant when the associated two-tailed p-value is less than 0.05. The mutual correlations among the controlled variables (e.g., between $\theta_{Bn}$ and magnetic compression ratio, or between upstream magnetic field strength and solar wind temperature) are weak ($|r| < 0.2$), indicating minimal multicollinearity and ensuring the robustness of the partial correlation analysis.

# 3 Statistical Study

## 3.1 Reflection Ratio

After data preprocessing, the remaining foreshock ion measurements are interpreted as solar wind ions reflected at the bow shock. The density ratio of these remaining ions to the solar wind density is therefore treated as the reflection ratio throughout this study.

We investigate the statistical dependence of the reflection ratio on a set of upstream and shock-related parameters, including the solar wind speed $V_{\mathrm{sw}}$, the magnetosonic Mach number $M_{\mathrm{ms}}$, the Alfvén Mach number $M_{\mathrm{A}}$, the shock normal angle $\theta_{Bn}$, and the bow shock magnetic compression ratio. The upstream parameters are obtained from OMNI data. The solar wind velocity is transformed from the GSE coordinates to the NIF coordinates, and the corresponding Mach numbers are calculated using the NIF velocity. Among these parameters, the highest correlations are found with the shock normal angle $\theta_{Bn}$ and the bow shock magnetic compression ratio using the downstream maximum (overshoot) magnetic field strength.

At the spin-averaged level, partial correlation analysis indicates a moderate negative correlation between the reflection ratio and $\theta_{Bn}$, with partial correlation coefficients on the order of $-0.3$ (Figure 2a). This trend is evident in the scatter distribution, where lower reflection ratios tend to be associated with more perpendicular shock geometries.

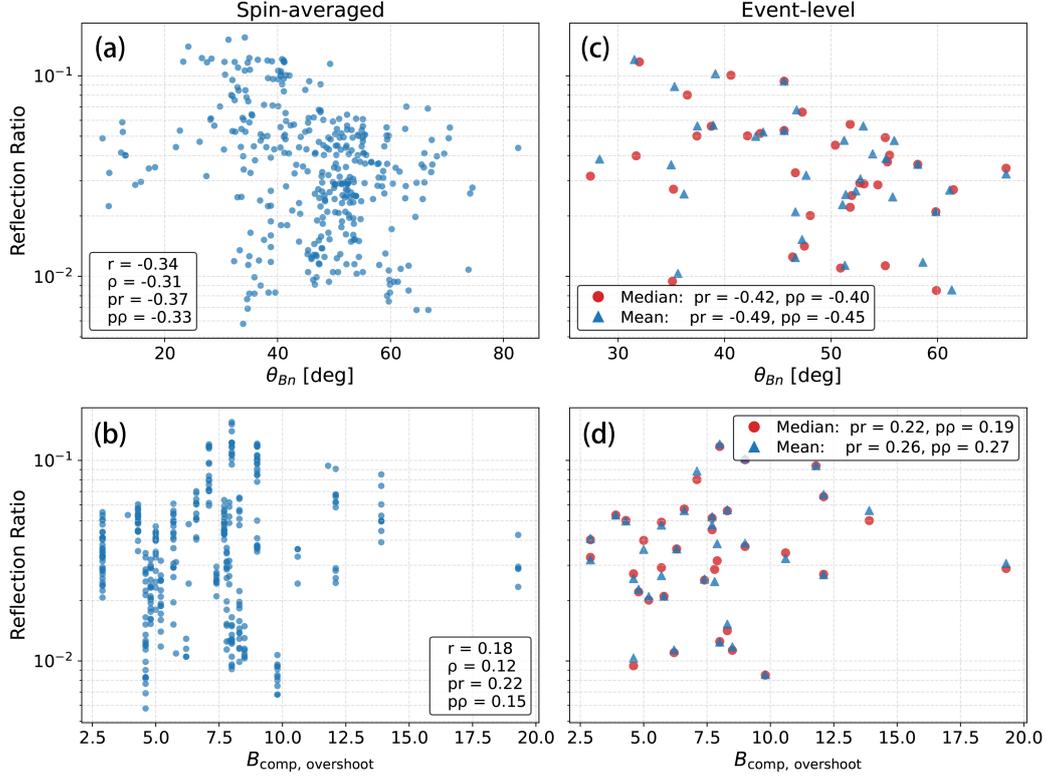

**Figure 2.** Scatter plots of the reflection ratio as a function of shock-related parameters. (a) and (b) show spin-averaged measurements, while (c) and (d) show event-level results. (a) Reflection ratio versus shock normal angle $\theta_{Bn}$. (b) Reflection ratio versus magnetic compression ratio $B_{\mathrm{comp,overshoot}}$, defined using the downstream maximum (overshoot) magnetic field strength. (c) Event-level reflection ratio versus $\theta_{Bn}$. (d) Event-level reflection ratio versus $B_{\mathrm{comp,overshoot}}$ (overshoot definition). In (c) and (d), red circles denote event medians and blue triangles denote event means. Pearson ($r$), Spearman ($\rho$), partial Pearson ($pr$), and partial Spearman ($p\rho$) correlation coefficients are indicated. Partial correlation coefficients are computed by controlling for the other shock parameter (i.e., $\theta_{Bn}$ or $B_{\mathrm{comp,overshoot}}$).

In contrast, the reflection ratio exhibits a weak positive correlation with the magnetic compression ratio, with partial correlation coefficients of approximately +0.2 (Figure 2b), indicating enhanced ion reflection under stronger magnetic field compression.

When the same analysis is performed at the event level, correlations with both $\theta_{Bn}$ (Figure 2c) and magnetic compression ratio (Figure 2d) become more pronounced. Compared to the spin-averaged results, the event-level partial correlation coefficients increase in magnitude for both parameters.

The dependence of the reflection ratio on the Alfvén and magnetosonic Mach numbers was also examined by replacing the magnetic compression ratio with the corresponding Mach number while retaining $\theta_{Bn}$ as the second control parameter. In this case, both the correlation and partial correlation coefficients decrease substantially, and no clear trend can be identified. The corresponding results are provided in the Supporting Information (Figures S1 and S2).

### 3.2 Velocity: Comparison with Reflection Models

We first examine the relationship between the energy derived from the observed reflected ion velocities and thermal contributions and the energy predicted by idealized reflection models. Both the adiabatic and specular reflection models describe changes in ion velocity without including additional thermalization. Since the upstream solar wind temperature at 1 au is typically on the order of $\sim$10 eV (e.g., Wilson III et al., 2018), its contribution to the total model energy is negligible compared to the bulk kinetic energy (which is around 1 keV for typical solar wind speeds). We therefore compare the observed ion kinetic plus thermal energy with the kinetic energy predicted by the reflection models.

In the adiabatic reflection framework, the reflected ion speed in the normal incidence frame (NIF) is given by

$$V_r^{\mathrm{NIF}} = V_u^{\mathrm{NIF}} \sqrt{1 + 4\tan^2\theta_{Bn}},$$

where $V_u^{\mathrm{NIF}}$ is the upstream solar wind speed in the NIF. The corresponding predicted ion energy therefore increases rapidly with increasing $\theta_{Bn}$.

Figures 3a and 3d compare the observed ion energy with that expected from the adiabatic reflection model. While a clear monotonic relationship is evident, the adiabatic reflection model systematically overestimates the observed energy. This behavior is reflected by relatively higher Spearman correlation coefficients compared to Pearson correlation coefficients, indicating that higher predicted energies are consistently associated with higher observed energies, but with a systematic mismatch in magnitude. In addition, the majority of data points lie above the $y = x$ line in Figures 3a and 3d, demonstrating that the model-predicted energies are generally larger than those inferred from observations. By contrast, the specular reflection model, by construction, does not include ion energization and therefore predicts reflected ion energies comparable to the incident solar wind kinetic energy. For typical solar wind speeds in our data set, this corresponds to energies around $\sim$ 1 keV, whereas the observed foreshock ion energies in Figures 3a are generally in the range of $\sim$ 2–10 keV. The specular prediction therefore substantially underestimates the observed ion energies and is not shown separately for clarity.

To further investigate the velocity-space characteristics of the reflected ions, we focus on the perpendicular motion. The gyro speed of the reflected ions is derived from the observed velocity distribution by extracting the perpendicular velocity component and removing the local $\mathbf{E} \times \mathbf{B}$ drift. When the gyro speed, together with the perpendicular thermal energy, is compared with the gyro energy predicted by the specular reflection model, a good correspondence is found (Figures 3b and 3e). This agreement in-dicates that the perpendicular energy content associated with the observed reflected ions is consistent with the specular reflection prediction.

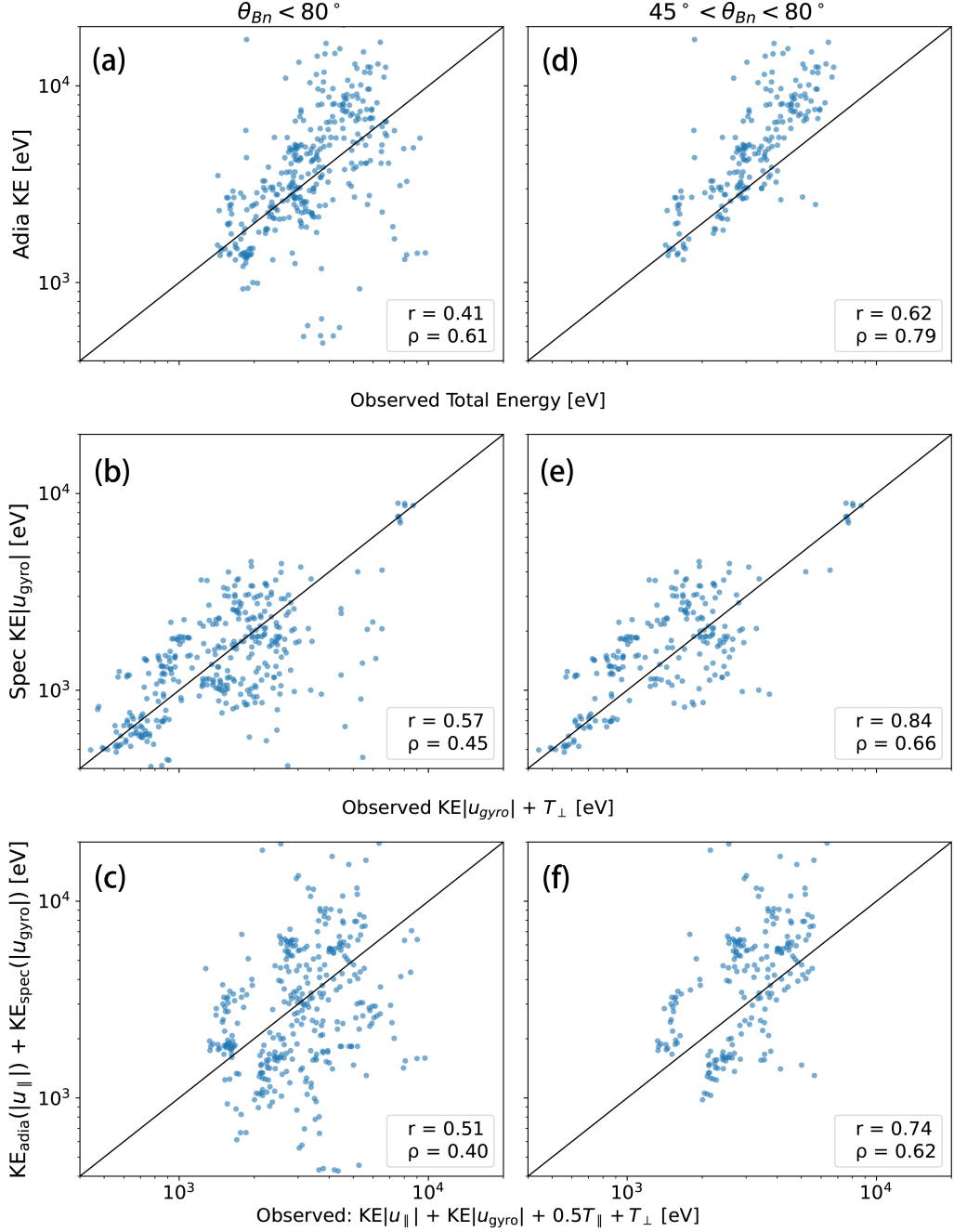

**Figure 3.** Comparison between modeled and observed reflected ion energies using spin-averaged data. The left column (a–c) includes data with $\theta_{Bn} < 80°$, while the right column (d–f) shows the subset of quasi-perpendicular shocks with $45° < \theta_{Bn} < 80°$. (a,d) Observed total ion energy (kinetic plus thermal) compared with the kinetic energy predicted by the adiabatic reflection model. (b,e) Observed perpendicular ion energy, defined as $KE(|u_{gyro}|) + T_\perp$, compared with the gyro kinetic energy predicted by the specular reflection model. (c,f) Combined model comparison, where the parallel velocity is taken from the adiabatic reflection model and the gyro velocity from the specular reflection model; the resulting modeled kinetic energy is compared with the associated observed ion energy (kinetic plus thermal). Pearson ($r$) and Spearman ($\rho$) correlation coefficients are indicated in each panel. The solid black line represents $y = x$, corresponding to perfect agreement between modeled and observed energies.

Motivated by the distinct physical assumptions underlying the two idealized models — magnetic mirroring primarily affecting the parallel motion and cross-shock potential governing the perpendicular (gyro) motion — we construct a combined representation in which the parallel velocity is taken from the adiabatic reflection model, while the gyro speed is taken from the specular reflection model. The energy associated with these modeled velocity components is then compared with the energy derived from the observed ion velocities and thermal contributions. As shown in Figures 3c and 3f, the correspondence between the modeled and observed energies is improved relative to the adiabatic reflection model alone, as evidenced by a higher Pearson correlation coefficient. In contrast, the Spearman correlation coefficient decreases compared to the adiabatic reflection model alone. This indicates that while the combined representation provides a better linear amplitude match, it does not further enhance the rank-order (monotonic) correspondence between modeled and observed energies. This result demonstrates that the combined representation provides a better quantitative match to the observed ion energy distribution.

We further examine the dependence of model–observation agreement on shock geometry. When restricting the analysis to quasi-perpendicular shocks ($45° < \theta_{Bn} < 80°$; Figures 3d–f), the correlation coefficients increase substantially compared to the full $\theta_{Bn} < 80°$ sample (Figures 3a–c). This enhancement is observed for all three model representations, indicating that the correspondence between modeled and observed energies is stronger in the quasi-perpendicular regime.

### 3.3 Temperature

We examine the statistical dependence of the reflected ion temperature on upstream and shock-related parameters using the same analysis framework as in the reflection ratio study. The tested parameters include the solar wind speed $V_{sw}$ in the NIF frame, the magnetosonic and Alfvén Mach numbers in the NIF frame, the shock normal angle $\theta_{Bn}$, the bow shock magnetic compression ratio, the local magnetic field strength measured by THEMIS FGM, the upstream IMF strength from OMNI, the upstream solar wind temperature $T_{sw}$, and the upstream plasma beta $\beta$ obtained from OMNI data. Among these parameters, the strongest correlations are found with the local magnetic field strength measured by THEMIS FGM and the upstream solar wind temperature. Partial correlation analysis indicates a moderate positive correlation with the local magnetic field strength, with partial correlation coefficients on the order of 0.5 as shown in Figure 4a, while the correlation with the upstream solar wind temperature is weaker, with partial correlation coefficients of approximately 0.2 as shown in Figure 4b.

To further investigate whether the observed temperature dependence on magnetic field strength is associated with magnetic field variability, we examine the relationship between the reflected ion temperature and var($\mathbf{B}$) in Figure 5a. In contrast to the correlations obtained using background field strength, the reflected ion temperature exhibits a substantially stronger correlation with var($\mathbf{B}$). The Pearson correlation coefficient reaches values close to 0.8, while the Spearman coefficient is moderately lower (on the order of 0.6), indicating a strong linear relationship with moderate rank-order correspondence. Similar correlation strengths are obtained when comparing spin-averaged and event-level median values.

In addition, we observe a positive correlation between var($\mathbf{B}$) and the total energy of reflected ions, with both Pearson and Spearman correlation coefficients of approximately $\sim 0.4$ (Figure 5b). This indicates a moderate correlation between magnetic field variance and reflected ion energization.

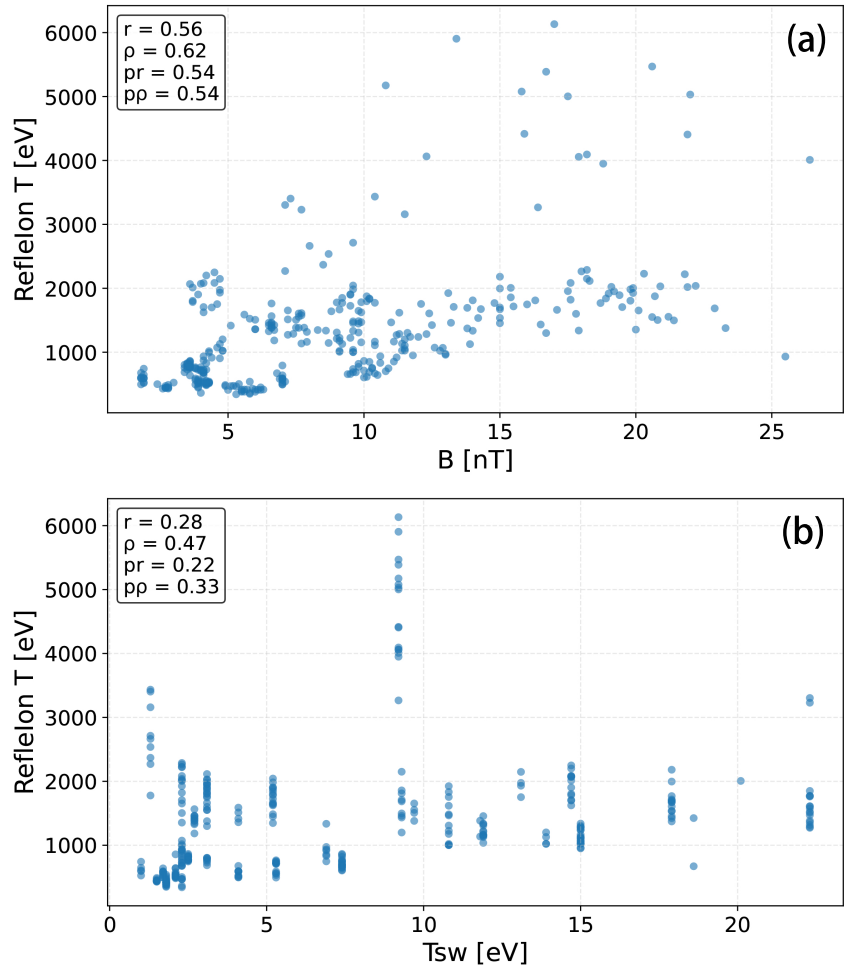

**Figure 4.** Scatter plots of the reflected ion temperature as a function of upstream parameters using spin-averaged data. (a) Reflected ion temperature versus upstream magnetic field strength $B$, measured by FGM. (b) Reflected ion temperature versus upstream solar wind temperature $T_{sw}$, obtained from OMNI data. Pearson ($r$), Spearman ($\rho$), partial Pearson ($pr$), and partial Spearman ($p\rho$) correlation coefficients are indicated. Partial correlation coefficients are computed by controlling for the alternate upstream parameter.

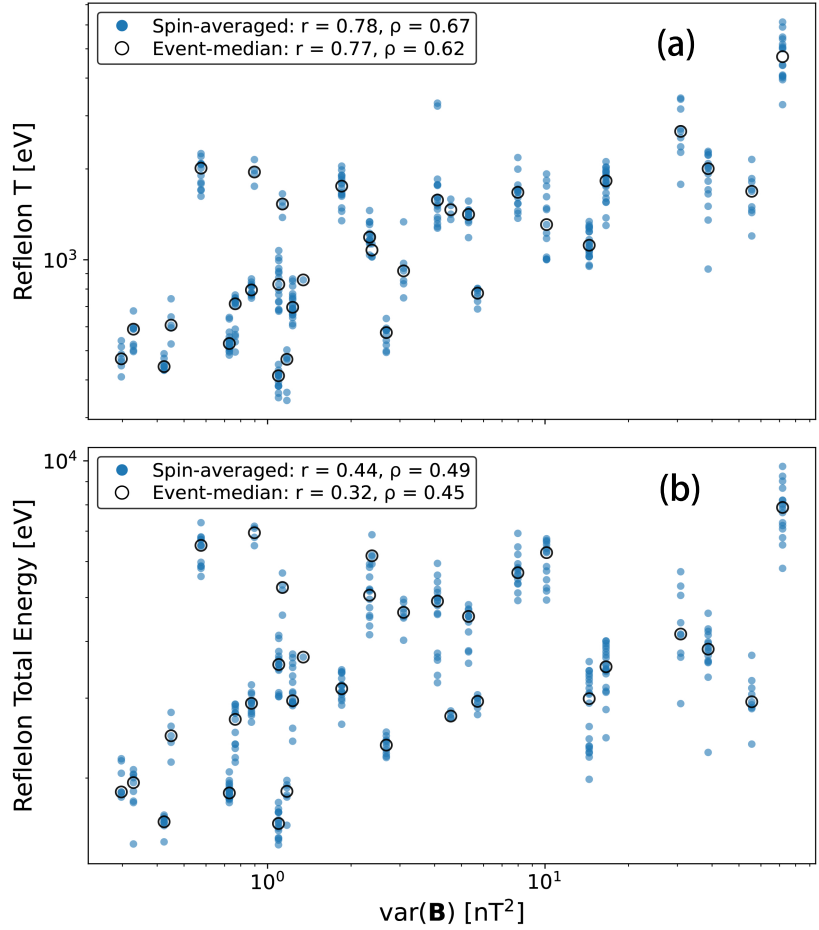

**Figure 5.** Scatter plots of (a) reflected ion temperature and (b) reflected ion total energy as a function of var(**B**). The quantity var(**B**) is computed for each event and is therefore constant for all spin-averaged measurements within the same event. Solid blue dots represent spin-averaged data, and open black circles denote event-level medians. Pearson ($r$) and Spearman ($\rho$) correlation coefficients are indicated in each panel.

## 4 Discussion

The statistical relationships identified in this study indicate that reflection ratio is primarily controlled by event-level properties. The stronger correlations obtained at the event level, compared to the spin-averaged cadence, suggest that parameters such as $\theta_{Bn}$ and magnetic compression ratio set the baseline reflection ratio, while short-timescale variability introduces additional modulation. Although Mach numbers were also examined, their partial correlations with the reflection ratio are systematically weaker than those associated with magnetic compression ratio. As shown in the Supporting Information (Figure S3), the Mach numbers exhibit a moderate to strong positive correlation with the magnetic compression ratio, with Pearson correlation coefficients lower than the corresponding Spearman coefficients, indicating a clear but not strictly linear relationship, consistent with previous results (e.g., Gedalin et al., 2021). This behavior suggests that Mach number may influence reflection ratio primarily through its regulation of shock magnetic compression rather than acting as an independent controlling parameter. One possible interpretation is that the bow shock typically operates in a high-Mach-number regime, where the role of Mach number in shock dissipation may be partially saturated.

The velocity analysis suggests that ion reflection behavior depends sensitively on shock geometry. Idealized reflection descriptions show improved applicability under quasi-perpendicular conditions, whereas their correspondence becomes less robust as the shock approaches the quasi-parallel regime. Together with the finding that the specular model better captures the perpendicular (gyro-associated) energy while the adiabatic model better reproduces the overall trend of the total energy, our results are consistent with scenarios in which multiple processes contribute to the observed reflected-ion energy budget, with their relative importance varying across shock conditions. The event-to-event variability in model–observation agreement further supports that no single idealized mechanism universally accounts for all cases.

For the reflected ion temperature, an initial positive correlation is observed with the local magnetic field strength. A stronger correlation, however, is found with var($\mathbf{B}$), which represents the variance of the magnetic field and is proportional to the magnetic fluctuation energy. As shown in the Supporting Information (Figure S4), the local magnetic field strength itself exhibits a strong positive correlation with var($\mathbf{B}$) (consistent with previous studies, e.g., Liu et al. (2025)), suggesting that the apparent dependence of ion temperature on magnetic field strength may arise indirectly through the associated increase in magnetic fluctuation energy. The strong correlation between ion temperature and var($\mathbf{B}$) therefore indicates that magnetic fluctuations play an important role in regulating foreshock ion thermal properties. The moderate correlation between var($\mathbf{B}$) and total reflected ion energy suggests that magnetic fluctuations may also influence the energization of reflected ions. One possible interpretation is based on wave–particle scattering: in the shock frame, magnetic fluctuations are convected toward the bow shock by the solar wind flow, while foreshock ions propagate away from the shock. Scattering interactions between the counter-streaming waves and ions can therefore lead to an increase in ion energy in the shock frame while preserving energy in the wave rest frame, as upstream part of the second-order Fermi process. However, detailed wave–particle analysis would be required to further quantify such effect.

Foreshock ions originate primarily from solar wind ion reflection at the bow shock and from magnetosheath ion leakage. The leakage component is influenced by both upstream parameters and downstream distributions (Liu, Angelopoulos, Zhang, et al., 2024). In this study, preprocessing steps were applied to reduce the influence of magnetosheath leakage and to preferentially retain locally reflected solar wind ions. Despite these efforts, the statistical results indicate that even the reflected solar wind ion population exhibits characteristics shaped by both large-scale upstream and shock properties and local variability, as evidenced by the behavior of the reflection ratio and ion temperature iden-tified in this study. These results suggest that reflected ion properties arise from the combined influence of upstream conditions, shock geometry, and local processes. Future modeling efforts should therefore account not only for global upstream and shock parameters but also for local processes and fluctuations that modulate ion reflection and heating.

Several limitations inherent to the observational methodology should be noted. First, the analysis is based on single-spacecraft bow shock crossings and assumes a constant shock normal direction and shock velocity during each event. In reality, the bow shock surface may exhibit spatial and temporal variability (e.g., Lowe & Burgess, 2003; Mazelle & Lembège, 2021). Such variability may introduce uncertainties in the derived shock parameters and in the transformation to the normal incidence frame. Second, the determination of the shock normal using the mixed-mode method relies on coplanarity assumptions. Although widely used, coplanarity conditions are not strictly satisfied under all shock configurations (e.g., Gosling et al., 1988), such as due to the presence of magnetosheath leakage (Liu, Angelopoulos, Vu, et al., 2024). Deviations from these assumptions may therefore introduce additional uncertainty to the estimated shock geometry. Third, the statistical correlations between shock-related parameters and reflected ion properties rely on the assumption that the observed upstream ions are locally generated at the shock crossing encountered by the spacecraft. In order for such correlations to be physically meaningful, the upstream conditions, the derived shock parameters, and the reflected ion population must correspond to the same shock region. While upstream solar wind parameters are generally relatively stable over the selected intervals, shock-related parameters may vary more significantly along the shock surface. Ions reflected at other portions of the shock may propagate along magnetic field lines and be sampled at the spacecraft location, potentially weakening the correspondence between locally derived shock parameters and the observed ion population. Although data preprocessing steps were implemented to reduce such effects, including short time interval (90 seconds) away from the shock crossing and the exclusion of solar wind discontinuities and transient intervals, nonlocal contributions cannot be entirely ruled out.

Future work should assess the robustness of the identified statistical relationships using hybrid or fully kinetic simulations. By testing whether the correlations between reflected ion properties and upstream or shock-related parameters arise self-consistently in numerical models, such studies can provide cross-validation of the observational conclusions.

## 5 Conclusions

In this study, we performed a statistical analysis of solar wind reflected ions near the Earth's bow shock using THEMIS observations from 59 well-defined shock crossings. The main results can be summarized in three aspects: reflection ratio, velocity characteristics, and temperature.

First, the reflection ratio exhibits a moderate negative correlation with the shock normal angle, $\theta_{Bn}$, and a weak positive correlation with the magnetic compression ratio defined using the overshoot magnetic field. These correlations become more pronounced at the event level compared to spin-averaged measurements, indicating that large-scale shock geometry and magnetic compression primarily regulate the baseline reflection ratio, while short-timescale variability introduces additional modulation.

Second, the observed reflected ion energies deviate systematically from idealized reflection models. The adiabatic reflection model reproduces the monotonic trend of the total ion energy but systematically overestimates its magnitude, whereas the specular reflection model captures the perpendicular (gyro-associated) energy component. A combined representation, in which the parallel velocity is taken from the adiabatic model and the gyro velocity from the specular model, improves the linear correspondence between modeled and observed energies. Model–observation agreement is strongest under quasi-perpendicular conditions.

Third, the reflected ion temperature shows a moderate positive correlation with the upstream magnetic field strength and a weaker correlation with the upstream solar wind temperature. More prominently, the reflected ion temperature exhibits a substantially stronger correlation with $var(B)$, indicating that magnetic fluctuation energy plays an important role in regulating the thermal properties of reflected ions.

These results provide observational constraints on the multi-parameter control of ion reflection and heating of the reflected ions at Earth's bow shock. Establishing such relationships improves the predictability of foreshock ion characteristics and contributes to a better understanding of foreshock-driven space weather phenomena in the near-Earth environment, while also providing guidance for future hybrid and fully kinetic simulations of foreshock ion populations.

### Acknowledgments

R.L. and K.Z. acknowledge support by the NSF grants AGS-2420710 and AGS-2247760. R.L. acknowledges support by the NASA FINESST Grant 80NSSC23K1633. The THEMIS mission is supported by NASA contract NAS5-02099.

### Conflict of Interest

The authors declare no conflicts of interest relevant to this study.

### Open Research

THEMIS data are provided by the THEMIS Science Support Center and are accessible at https://themis.ssl.berkeley.edu/data/themis/. The specific time intervals analyzed in this study are described in the main text and figures. Solar wind and interplanetary magnetic field parameters were obtained from the OMNI data set, available at https://omniweb.gsfc.nasa.gov/. Data analysis was performed using the SPEDAS software package, which can be found at https://spedas.org.